\documentclass[12pt,a4paper]{article}
\usepackage[T1]{fontenc}
\usepackage[utf8]{inputenc}
\usepackage{authblk}
\expandafter\let\csname equation*\endcsname=\relax 
\expandafter\let\csname endequation*\endcsname=\relax 
\usepackage{amsmath,amssymb,amsthm} 
\usepackage{graphicx}
\usepackage{lineno,hyperref}
\usepackage{wrapfig}
\usepackage{amsmath}
\usepackage{graphicx}
\usepackage{epstopdf}
\usepackage{multirow}
\usepackage{bigstrut}
\usepackage{wrapfig}
\title{A low cost hybrid detection system of high energy air showers}
\author[a]{A.G. Tsirigotis}
\author[a]{A. Leisos}
\author[b]{S. Nonis}
\author[a]{M. Petropoulos}
\author[a]{G. Georgis}
\author[b]{K. Papageorgiou}
\author[b]{I. Gkialas}
\author[c]{I. Manthos}
\author[c]{S.E. Tzamarias}
\affil[a]{Physics Laboratory, School of Science \& Technology, Hellenic Open University, Patras, Greece}
\affil[b]{Department of Financial and Management Engineering, University of the Aegean, Chios, Greece}
\affil[c]{Department of Physics, Aristotle University of Thessaloniki, Thessaloniki, Greece}

\date{}

\begin{document}
\maketitle
\begin{abstract}
We report on the design and the expected performance of a low cost hybrid detection system suitable for operation as an autonomous unit in strong electromagnetic noise environments. The system consists of three particle detectors (scintillator modules) and one or more RF antennas. The particle detector units are used to detect air showers and to supply the trigger to the RF Data acquisition electronics. The hardware of the detector as well as the expected performance  in detecting and reconstructing the angular direction for the shower axis is presented. Calibration data are used to trim the simulation parameters and to investigate the response to high energy ($E>10^{15} eV$) extensive air showers.
\end{abstract}
\section{Introduction}
The cosmic ray showers are traditionally detected using large arrays of particle detectors. In the sixties, a new, complementary method of detecting cosmic ray showers was proposed, using the RF signature of a cosmic ray event. This mechanism \cite{Askaryan_radio_emission}, exploits the electromagnetic signal emitted by the electromagnetic interactions of the shower charged particles.
The advantage of this method of detection is the large duty cycle, being of the order of 95\%, compared to other methods such as fluorescence and Cherenkov. However the relative weak RF signals from air showers suggests the application of this technique to radio quite environments far away from cities and man made electromagnetic noise. However, the RF signals of high energy showers can also be detected in strong electromagnetic noise environments by combining the experimental information from particle detectors and radio antennas as it was demonstrated mainly by the LOPES experiment \cite{lopes}. 

Beyond large scale detection systems, also small scale hybrid detection systems are capable for detecting the RF signals of high energy showers. Astroneu\cite{astroneu,helycon,bourlis} is such a detector. 
The current Astroneu array consists of nine large Scintillator Detector Modules (SDM) and six bipolar RF Antennas (RFA), installed at the Hellenic Open University (HOU) campus in Patras. During the first pilot run (2014-2017) three autonomous stations were operated, each comprising of three SDM and one RFA .
More than 600,000 showers recorded in a operating period of 1.5 years were used to study the efficiency and resolution of the array in single station and double station operation mode. A large fraction of events detected simultaneously by two distant stations are caused by high energy EAS that can also be detected by the RF antennas. Selection criteria applied to the digitized pulses confirmed  the RF signature of Cosmic events \cite{rfhelycon}.

In Section \ref{design} the design of a low cost system hybrid EAS detector is described comprised of 3 small SDMs and a number of RFAs (hereafter called 3SDM-nRF). Section \ref{simulation} describes the simulation of the system, Section \ref{showers} presents the system's responce to showers in comparison with an Astroneu station, and Section \ref{performance} shows the performance of the system in shower detection and reconstruction using the signal from the RFAs.
\section{The Design of the 3SDM-nRF system}
\label{design}
\begin{figure}[h]
\includegraphics[width=0.99\linewidth]{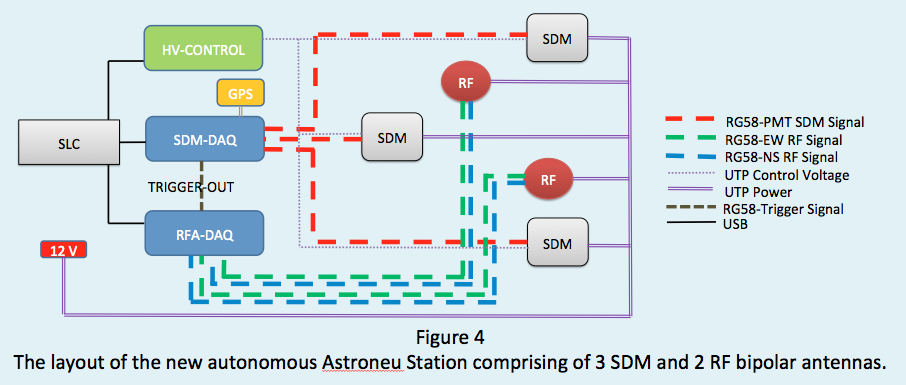}
\centering
\caption{The layout of the new autonomous small scale Astroneu Station comprising of 3 SDM and 2 RF bipolar antennas.}
\label{fig:sketch}
\end{figure}

The Physics Laboratory of HOU designed a low cost semi autonomous station for the extension of the Astroneu array with more particle detectors and RF antennas. The sketch layout of the station is shown in Figure \ref{fig:sketch}.
The proposed station comprises of 3 SDM and 2 RF antennas. The SDM pulses are acquired by the SDM-DAQ which applies a 3-fold majority logic, provides the timing of the crossing of the rising edge of the waveform with an adjustable voltage threshold, the Time over Threshold (ToT), timestamps the event using GPS, and sends a trigger-out signal to the RFA-DAQ. The functionality of the SDM-DAQ is provided by the QuarkNet\footnote{Due to the limited availability of the QuarkNet card, a system of equivalent functionality is designed by the Astroparticle Physics group of HOU. This system will consist of a high sampling rate USB oscilloscope, a majority logic coincidence electronics board, and a GPS unit capable to timestamp the trigger output pulse of the coincidence board.} card used also in Astroneu and described in \cite{astroneu}. The HV-Control module is a Digital to Analog Converter (DAC) providing the input signal for a dc-dc converter inside the SDM, that powers a Silicon Photomultiplier (SiPM). The RFA-DAQ is a 200 MHz analog bandwidth PC-based oscilloscope with 4 channels and external trigger (Hantek-DSO3204A). The waveforms of each dipole (in EW and NS direction) of the RF antennas (4 in total) are fully digitized with a sampling rate of 250 MSa/s. The analog bandwidth and sampling rate of the RFA-DAQ is adequate for the RF signals of interest in the region 20-80 MHz \cite{rfnonis1}. The SDM-DAQ, HV-Control and RFA-DAQ are connected and operated by the Station Local Computer (SLC) through the USB interface.
\subsection{Particle Detectors}
\begin{wrapfigure}{R}{0.5\textwidth}
  \centering
  \includegraphics[width=0.48\textwidth]{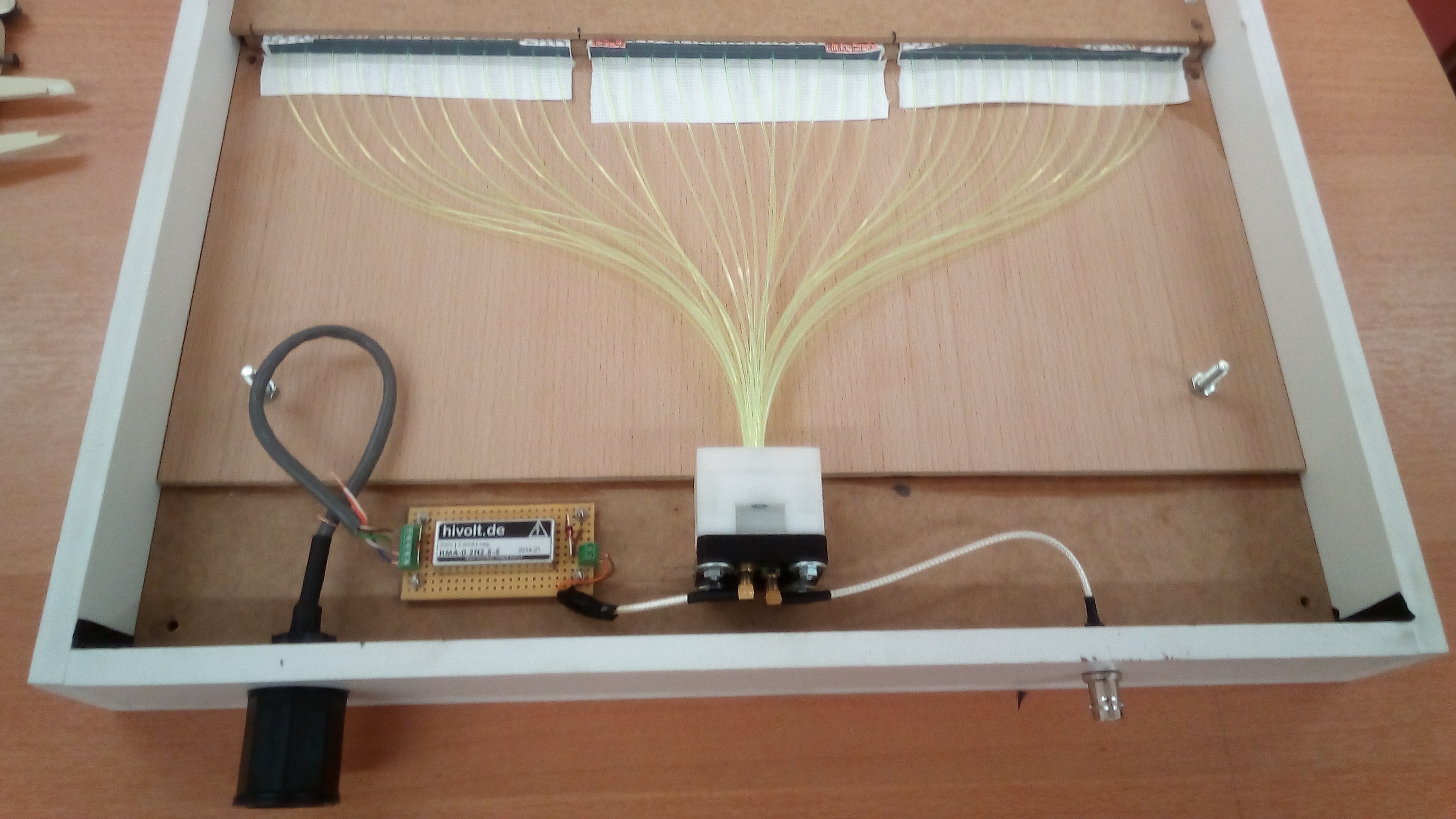}
  \caption{The SDM before light proofing and closing it. In lower left is the dc-dc converter providing the needed voltage to the SiPM in the lower center. The edge of the scintillation tiles and the WLS fibers are also shown.}
  \label{fig:SDM}
\end{wrapfigure}
The SDM active area of the prototype is $36 \times 40~cm^2$ and is made by two layers of scintillation tiles (12 tiles per layer) wrapped with reflective paper (Tyvek). Wavelength shifting (WLS) fibers (36 per SDM) are used to guide the light to a Silicon Photomultiplier (SiPM) operated at $30-33~V$. The overall dimension of  each particle detector unit is  $67 \times 42 \times 7~cm^3$ and weights 6 kg (Figure \ref{fig:SDM}). 

The SiPM has a photosensitive area of $6 \times 6~mm^2$ and is either the KETEK~PM6650-EB or the SensL~MICROFJ-SMTPA-60035-GEVB. The operation voltage of the SiPM is in the range $30-33~V$ and is provided by a dc-dc converter (hivolt.de HMA-0.2N2.5-5), which in turn is controlled by the HV-Control module, a DAC (Lucid Control AO4) connected through the USB interface with the SLC. More information about the design and the construction procedures of the SDM can be found in the reference \cite{miccos}.
\subsection{RF antenna}
One of the main components of the 3SDM-2RF system is the Low Noise Amplifier (LNA). The Physics Laboratory of HOU designed a low cost LNA that is placed near the RF dipole. The LNA has an amplification of 30~dB, four single lead inputs for each of the 4 poles of the 2 dipoles (EW and NS directions) and outputs through 2 $BNC/50 \Omega$ connectors. The LNA is optimized for each Antenna geometry to work as a band pass filter in the frequency band 20-80 MHz.

\begin{figure}[h]
\centering
\includegraphics[width=0.99\textwidth]{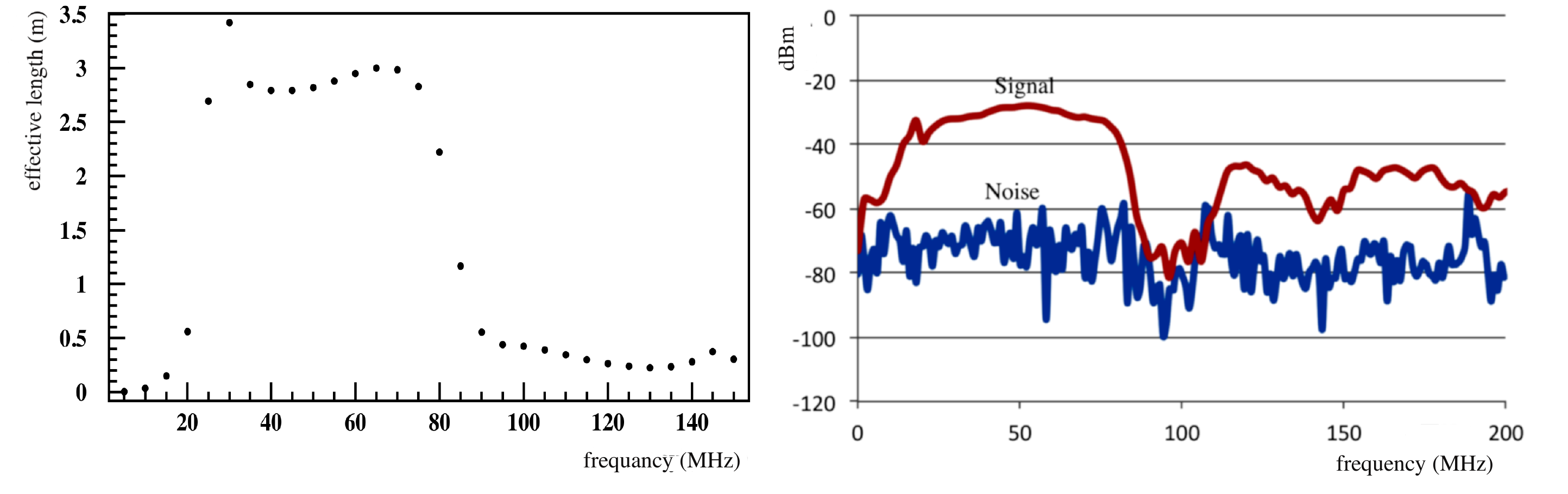}
\caption{Left: The effective length of a Codalema type butterfly antenna and an LNA optimized accordingly. Right: The noise spectrum at the installation site of a Codalema type butterfly antenna.}
\label{fig:antenna}
\end{figure}

Several RF antenna dipole geometries are under consideration; from a simple rod to Codalema butterfly antenna \cite{codalema},  or bowtie antenna with rectangular ends. The antenna characteristics are calculated by the 4NEC2 software \cite{4nec2}. For the Codalema type butterfly antenna and an LNA optimized accordingly, the effective length\footnote{We define the effective length as the average over all RF incident directions of the ratio of the voltage produced at the leads of the antenna over the incident magnitude of the electric field. This effective length depends not only on the antenna geometry but also on the input impedance of the LNA.} as a function of the frequency is presented in Figure \ref{fig:antenna}:Left.  At the same frequency regime the noise spectrum at the installation site was measured and it is shown in comparison with a typical cosmic event spectrum (Figure \ref{fig:antenna}:Right).

The methodology used for the design of the LNA is based on the 4NEC2 software. For each antenna geometry, its dimensions are adjusted so that its resonant frequency is in the middle of the 20-80MHz band, i.e. 50MHz. The impedance of the antenna is calculated and the input impedance of the LNA is optimized so that it best matches the antenna impedance, while at the same time the LNA filters out frequencies below 20MHz or above 80MHz. These frequencies are full of noise due to strong AM and FM radio broadcasts in urban environments. Moreover, the LNA input impedance is adjusted so that the magnitude of the antenna effective length do not vary significantly in the frequency band 20-80MHz.
\section{Simulation Tuning and Calibration }
\label{simulation}
The calibration of the particle detector units include the determination of the detector response to Minimum Ionizing Particles (MIPs),  the synchronization of the detectors and the determination of timing offsets due to the pulse size (slewing), as well as the spread of the pulses arrival time distribution.
\begin{figure}[h]
\includegraphics[width=0.49\linewidth]{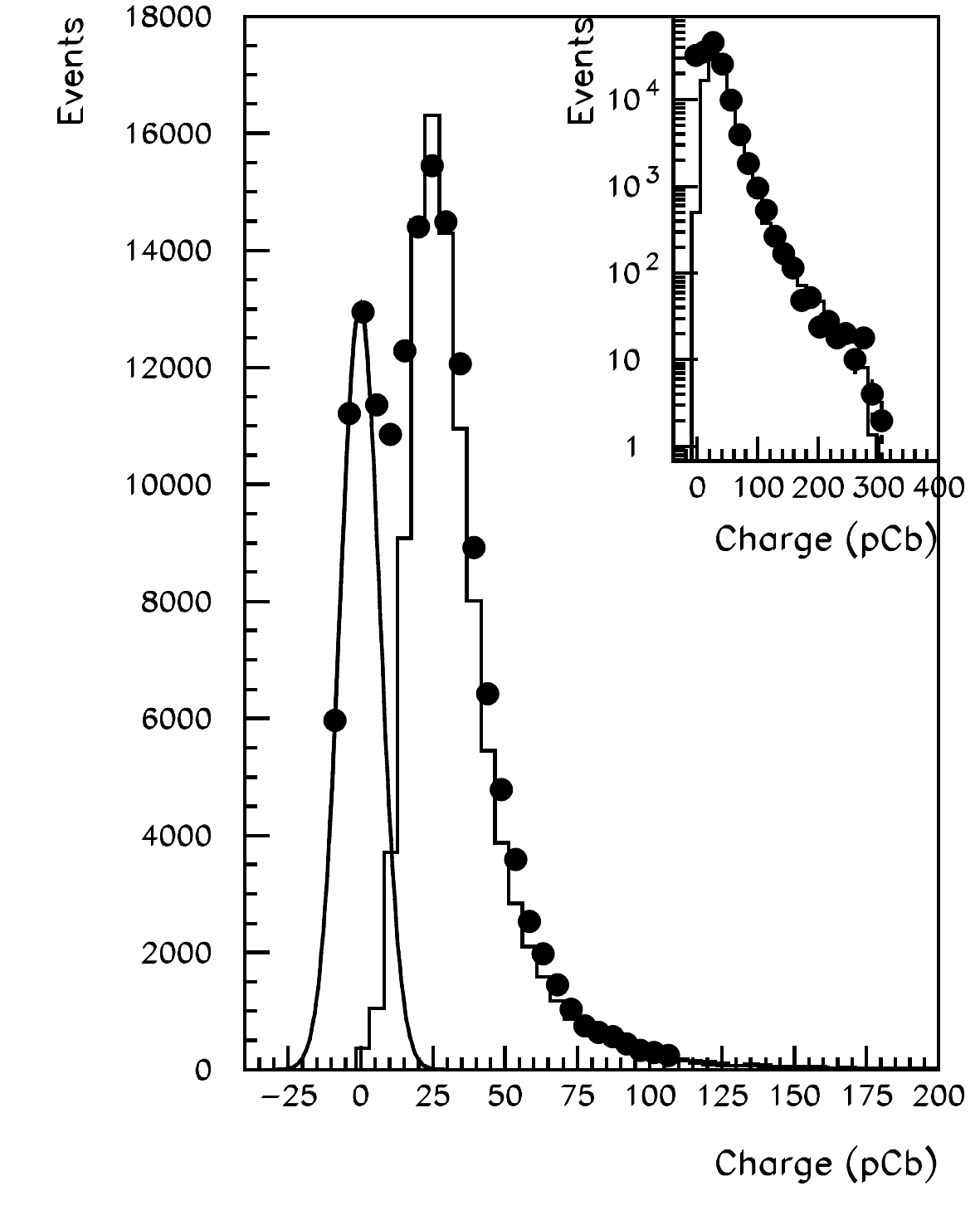}
\includegraphics[width=0.49\linewidth]{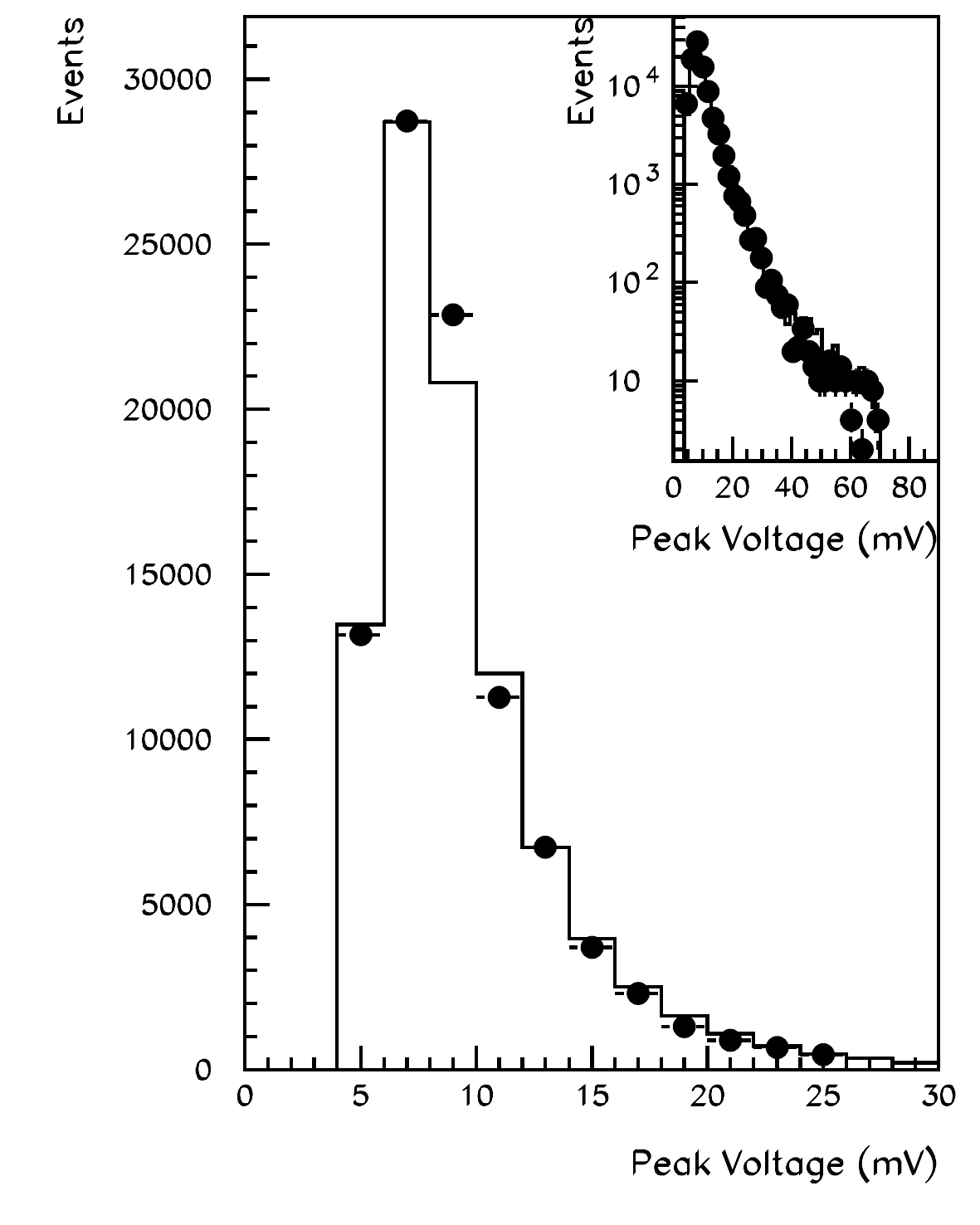}
\centering
\caption{The charge (left plots) and pulse height (right plots) distributions for experimental data (dots) and simulation (histogram). The insert plots are the distributions in logarithmic scale. The peak for data at zero charge corresponds to noise where the measured SDM had no significant pulse.}
\label{fig:simtun}
\end{figure}
 In order to study these effects and tune simulation parameters, three SDMs were arranged on top of each other. Events where at least two of them had a pulse above a predefined threshold (5~mV) where used to produce the calibration information. The time when the rising edge of the pulse crossed the threshold of 5~mV, as well as the corresponding ToT were recorded.
 The charge collection efficiency and the timing parameters of the scintillation and WLS processes were determine using methods described in reference \cite{calib}. In the plots of Figure \ref{fig:simtun} the response of the SDMs to MIPs in terms of charge distribution and peak voltage distribution are presented. For these plots two specific SDMs of the stack were chosen to provide a coincidence trigger, while the pulses of the third one were digitized.

\begin{wrapfigure}{R}{0.5\textwidth}
\centering
\includegraphics[width=0.48\textwidth]{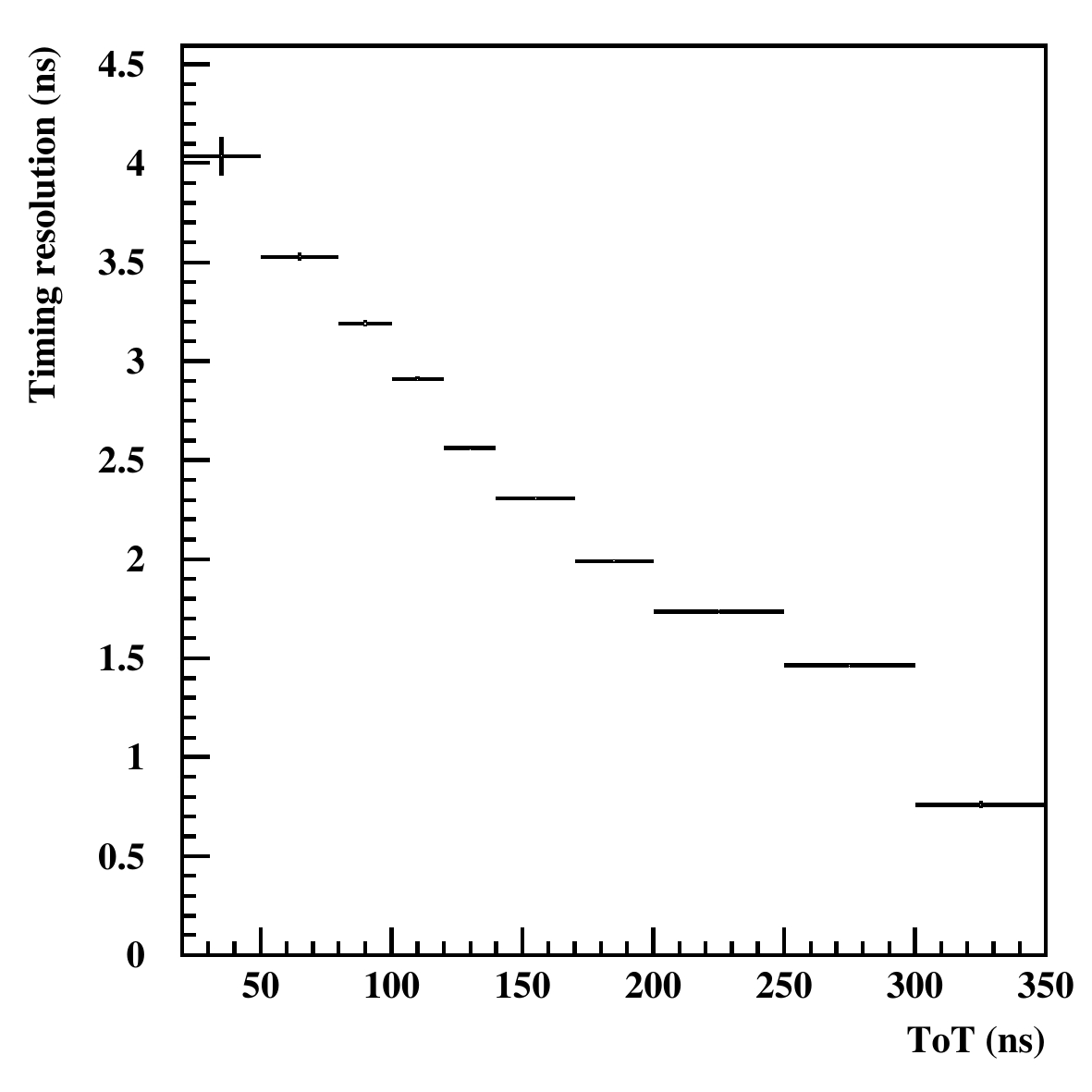}
\caption{The pulse arrival time resolution of the SDMs as a function of the ToT of the pulse.}
\label{fig:rms}
\end{wrapfigure}

For the timing characteristics of the pulses we estimated the slewing correction versus the ToT of the pulse, by comparing the pulse arrival time differences of two SDMs. The dependence of the slewing correction on the pulse ToT were found to follow the relation:
\begin{eqnarray}
  {t_{cor}=31\cdot e^{-0.0075 \cdot ToT}~(ns)~,~ToT<120~ns}\nonumber \\
  {t_{cor}=27.5\cdot e^{-0.0065 \cdot ToT}~(ns)~,~ToT>120~ns}\nonumber
\end{eqnarray}
The pulse arrival time resolution of the SDMs is shown in Figure \ref{fig:rms}. This resolution is estimated by comparing the corrected for slewing pulse arrival times of the SDMs for a number of ToT regions.

The simulation of the 3SDM-2RF system comprises of several steps. The first step is the production of the high energy showers, capable of activating all 3 SDMs, while at the rime time have enough energy to give a considerable RF signal to the dipole antennas. The EAS fulfilling these two prerequisites are caused by primary cosmic rays of energy above $10^{17}~eV$. Using CORSIKA \cite{corsika}, showers with an equivalent life time of 350000 hours\footnote{For each simulated shower, its impact point was randomly distributed 1890 times in a radius of 420~m around the detector array center.}, in the energy range $10^{15} - 10^{18}~eV$, were simulated with primary relative abundances and spectral index according to the latest measurements \cite{abundance}.

The second step is the simulation of the response of the SDMs to the secondary particles of the shower, as they were recorded by CORSIKA. The MC package used for this simulation is the same one used for the Astroneu array and is described elsewhere \cite{calib}. This package describes the number and timing profile of the scintillation photons that hit the SiPM photosensitive area for each secondary particle of the EAS incident on the SDM. This information depends on the type of particle and the incident position on the SDM.

The third and last step is the production of the simulated pulses, using also the experimental information of the response of the SDMs to MIPs acquired during the calibration procedure described above. The results of the calibration procedure where also used during the analysis of the produced simulation information, to correct the timing of the pulses for the slewing effect. 
\section{Response to showers}
\label{showers}
\begin{figure}[h]
\centering
\includegraphics[width=0.8\textwidth]{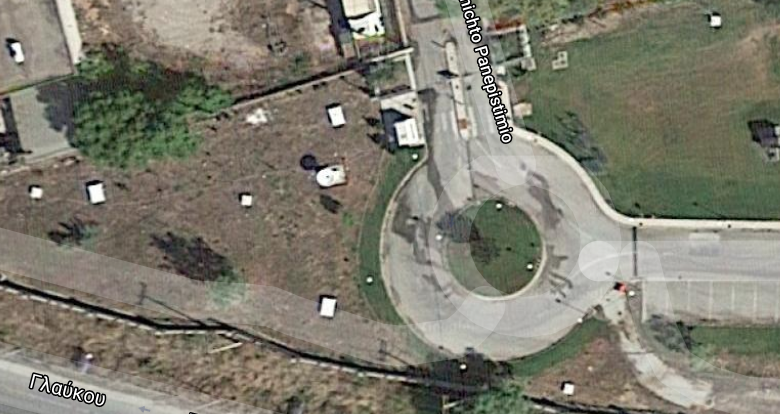}
\caption[width=0.4\textwidth]{Aerial view of the station A of Astroneu array. The rectangular white spots are the 3 Astroneu SDMs, while the smaller white squares are the 4 Codalema RF antennas. The SDMs define an isosceles triangle with base 22~m and large side 29~m.}
\label{fig:map}
\end{figure}

The triggering efficiency of the particle detectors was tested against one of the stations of the Astroneu array.
The SDMs of the 3SDM-2RF were positioned under the large Scintillator Detector Modules of station-A were 4 RF CODALEMA butterfly antennas are triggered upon a coincidence of the 3 detector units of station A (Figure \ref{fig:map}).
The 3SDM-2RF detection system under test was operated independently and the association between triggers was performed offline by checking the corresponding GPS time stamps.    
\begin{figure}[h]
\includegraphics[width=0.49\linewidth]{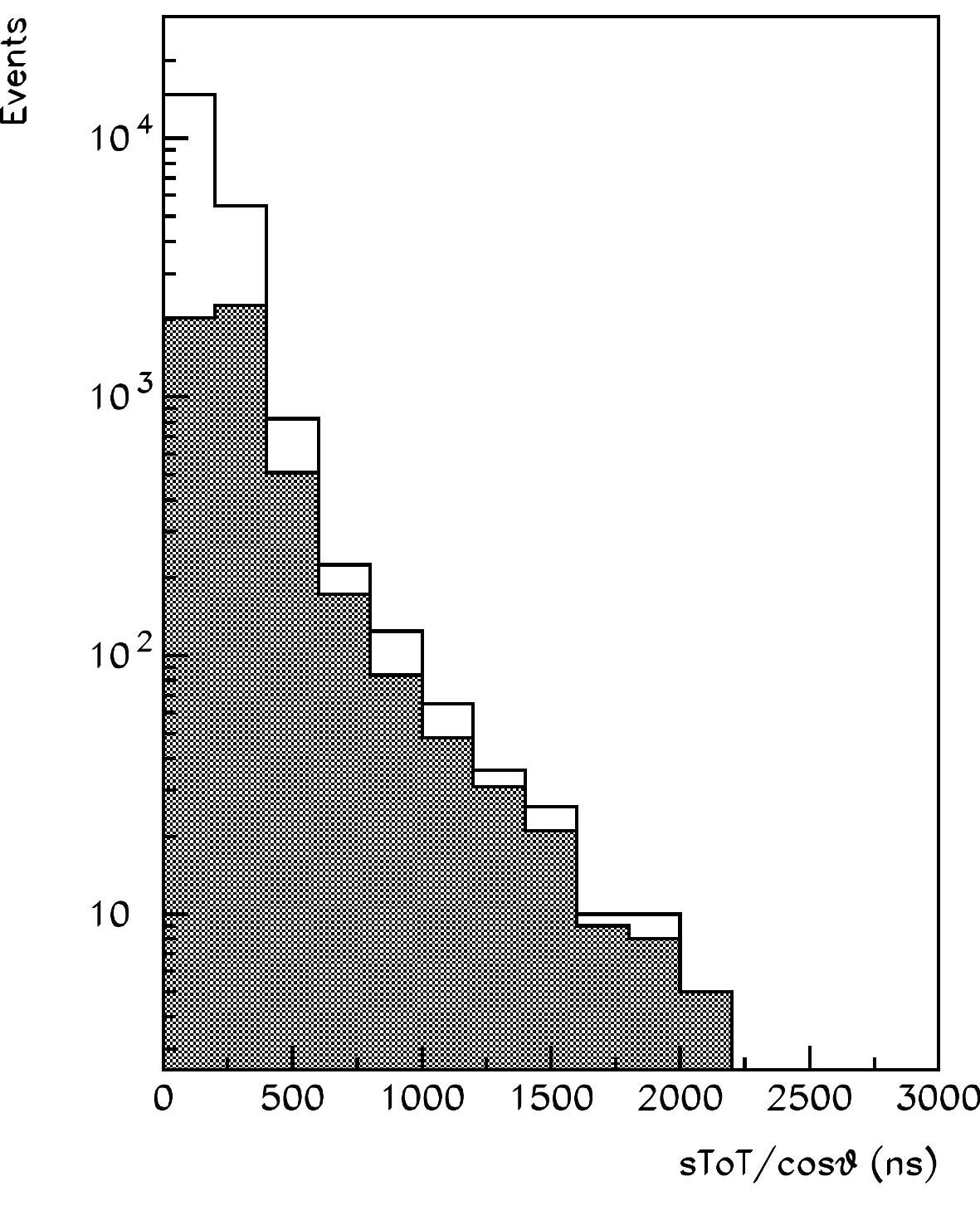}
\includegraphics[width=0.49\linewidth]{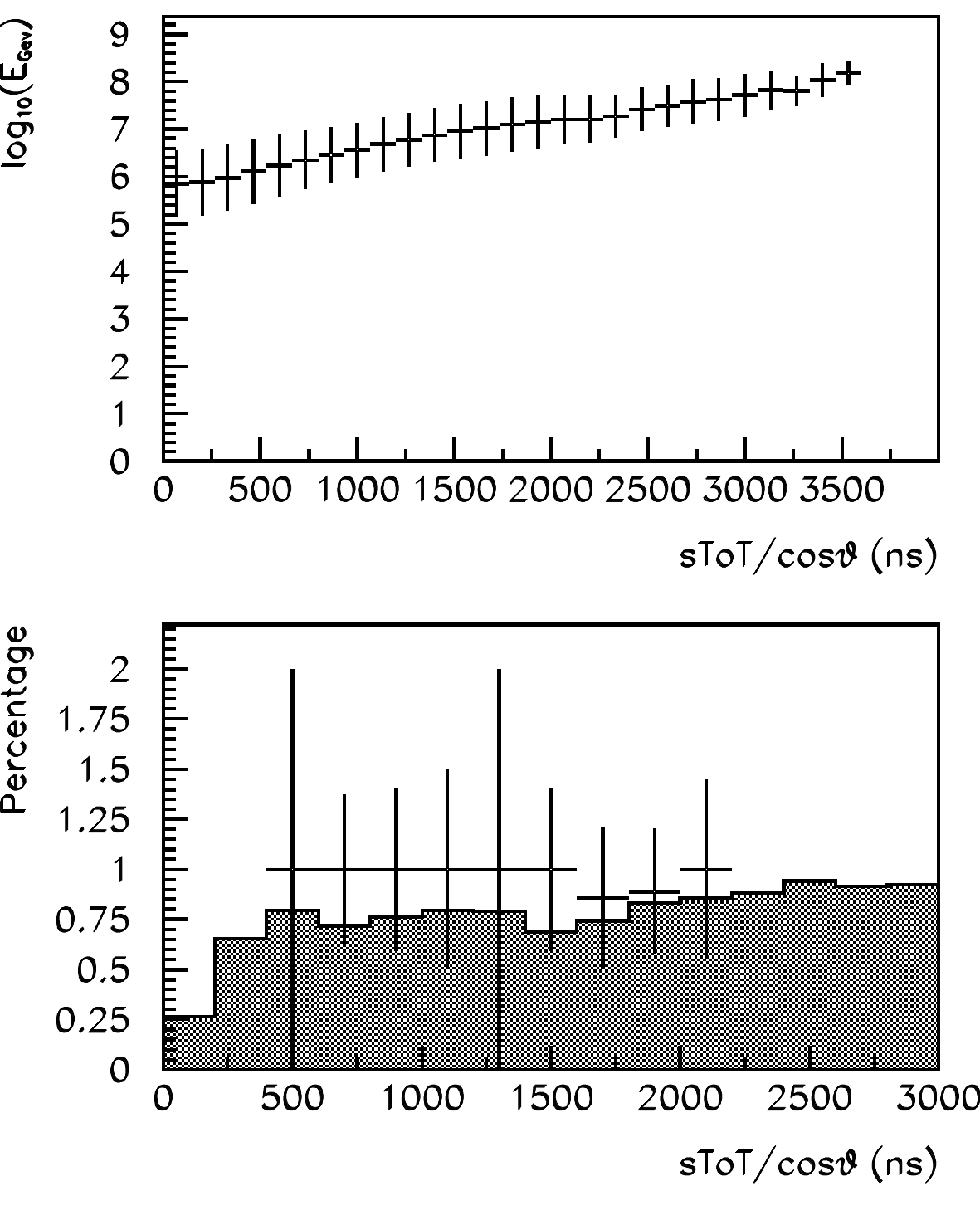}
\centering
\caption{Left: Distribution of showers detected by the Astroneu station (white histogram), or detected by both Astroneu and 3SDM-2RF station (shaded histogram) (see text also). Bottom right: The trigger efficiency of the 3SDM-2RF station for data (points with error bars) and MC (shaded histogram) with respect to the Astroneu station for showers that are detected by the Astroneu station and also register RF signal on the Astroneu antennas. Top right: Relation of the simuated energy versus the sum of ToTs for the detected events of 3SDM-2RF station.}
\label{fig:trigger}
\end{figure}

Figure \ref{fig:trigger}:Left shows the distribution of the $SToT/cos(\theta)$ (the sum of all the ToTs over the projection of a unit area perpendicular to the shower direction) of the Astroneu SDMs for all the reconstructed showers (white histogram) in comparison with the corresponding distribution for showers that were also detected by the 3SDM-2RF station (shaded histogram). From this plot it is evident that the relative shower detection efficiency of the 3SDM-2RF detector modules in respect to the Astroneu SDMs is very good, and approaches 100\% as the shower's energy increases. Moreover, for the most interesting cases of showers, where the Astroneu RF antennas have also signal correlated with the Astroneu SDMs, the 3SDM-2RF system's efficiency is even better as it shown in Figure \ref{fig:trigger}:Bottom right. The measured efficiency is in a very good agreement with the simulation prediction (shaded histogram).

These performance results show that a 3SDM-nRF system with 3 small SDMs and a number of RFAs can have similar performance with an Astroneu station having 3 large SDMs.
\section{Expected Performance}
\label{performance}
\begin{figure}[t]
\includegraphics[width=0.32\textwidth]{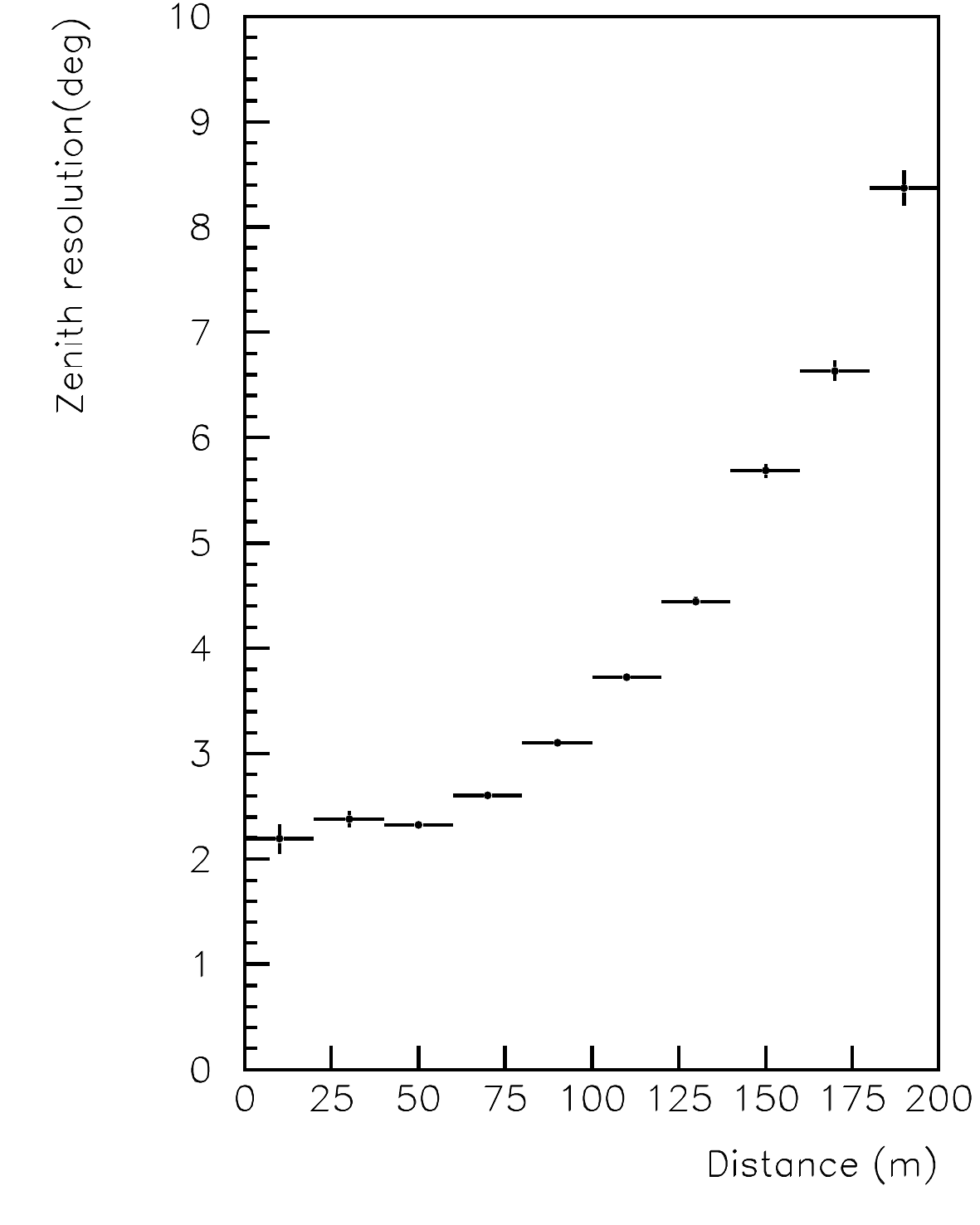}
\includegraphics[width=0.32\textwidth]{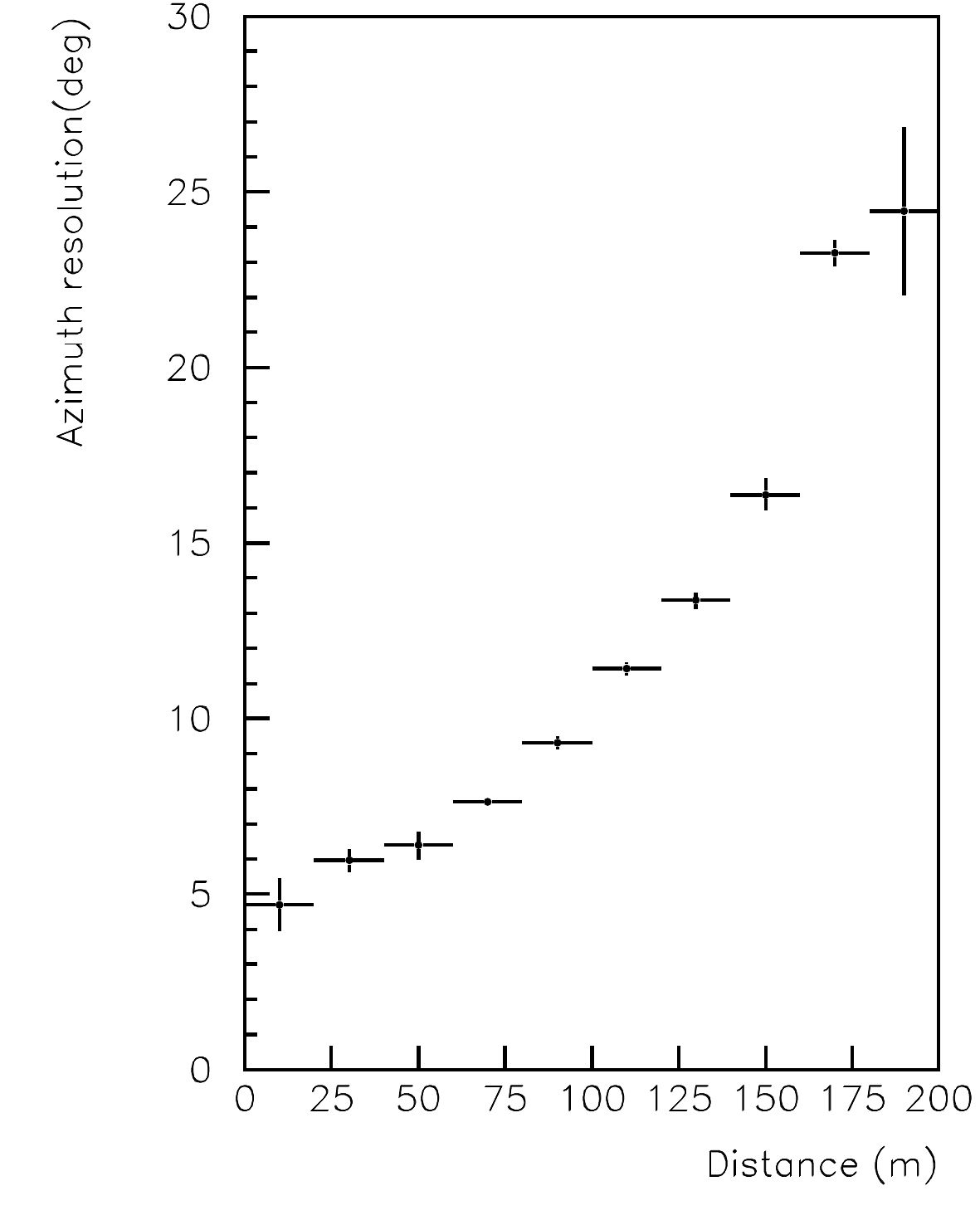}
\includegraphics[width=0.32\textwidth]{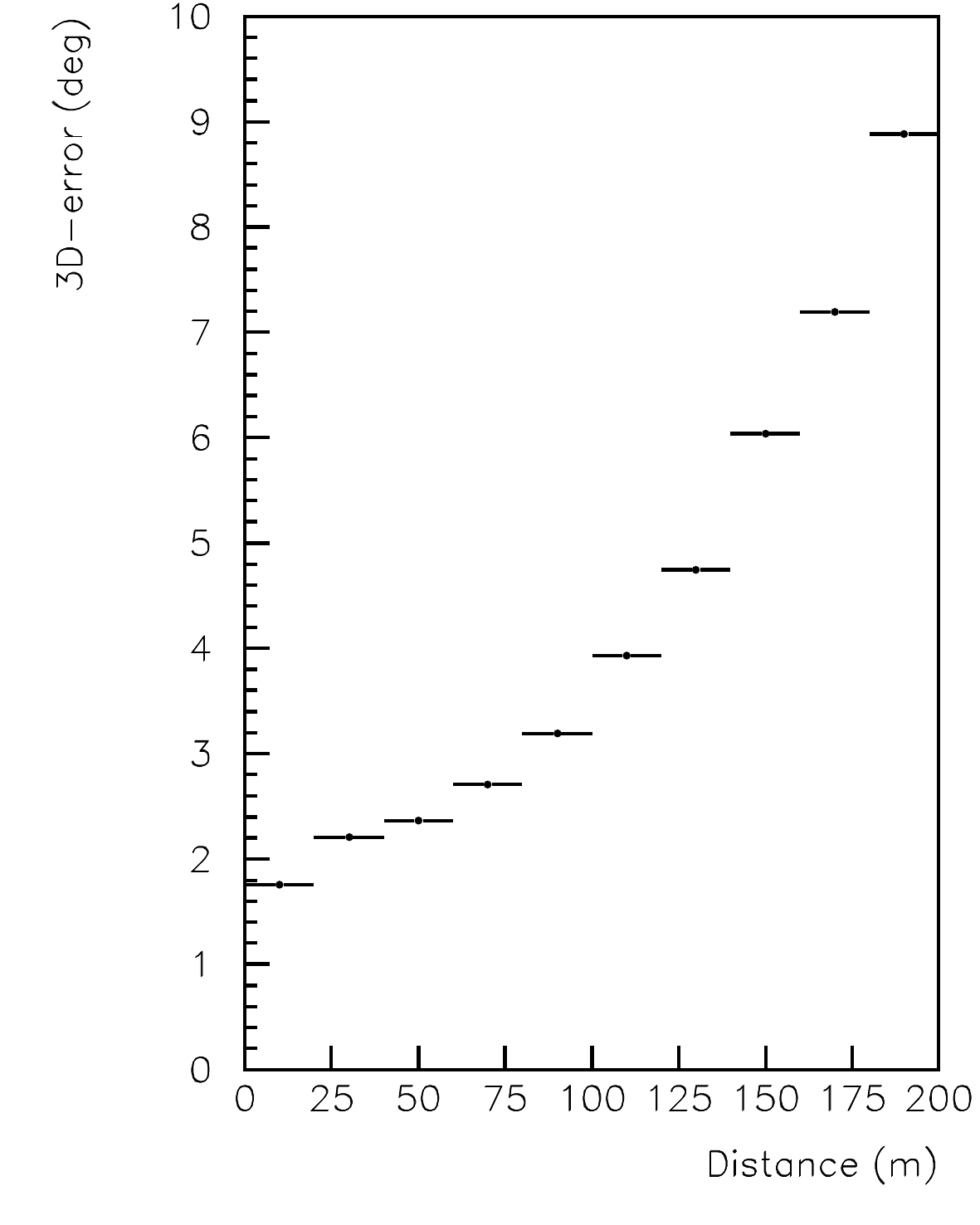}
\centering
\caption{Reconstruction resolution of the shower axis direction using only the RF signal of the four antennas of Station A, as a function of the distance of the Station A center to the shower impact point. From left to right are: zenith resolution, azimuth resolution, and 3D-angle distance of the simulated and reconstructed shower axis directions.}
\label{fig:angle}
\end{figure}
\begin{figure}[h]
\centering
\includegraphics[width=0.6\textwidth]{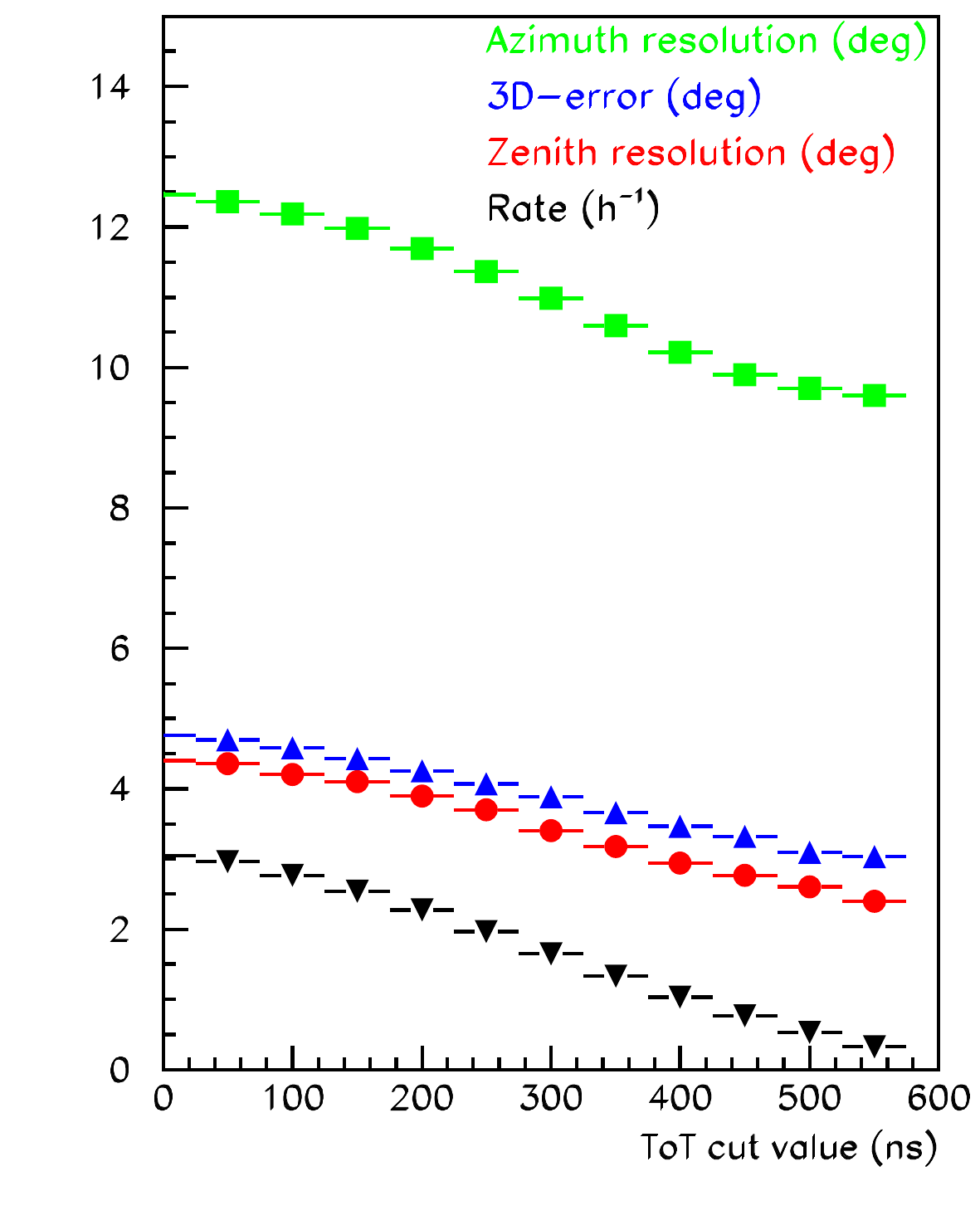}
\caption{Reconstruction resolution, as of Figure \ref{fig:angle}, versus the cut value on the sum of ToTs of the Astroneu SDMs (see also text).}
\label{fig:effic}
\end{figure}

In order to quantify the performance of the detection system a large number of showers were produced in the energy range $10^{15}-10^{18}~eV$. The detected showers were reconstructed by the spectrum response of the antennas as it is described in \cite{rfnonis2}. The procedure involves a minimization procedure where the shape of the power spectrum of the Fourier transformed RF pulse is compared with the simulation prediction. The resolution of the reconstructed shower-axis direction is presented in Figure \ref{fig:angle}  as a function of the distance between the shower axis impact point and the center of station A. The zenith and azimuth error are expressed as the square root of the variance between the true primary value and the reconstructed value, while the error on the angle between the primary direction and the reconstructed shower axis direction (3D-angle) as the median of the corresponding distribution. 
Due to the curvature of the shower front, the angular resolution depends on the proximity of the shower axis impact point to the center of the array. By applying a cut on the charge collected by each SDM detector, we select showers that are closer to the station center (higher cut value). In Figure \ref{fig:effic} is shown the resolution on the zenith, azimuth and 3D-angle as a function of the cut value. As it is expected higher cut values result in higher resolution but lower reconstruction rate which is also shown on the same plot.
\section{Conclusions}
A low cost hybrid detector system for high energy shower was designed, using scintillation detector modules for the particle component of the showers and RF antennas for the electromagnetic component. This system's performance study was done using as reference a station of a Astroneu array, with well known response to air showers. The results of this study show a very good performance of the low cost hybrid detector in detecting showers. Several of such systems, with a cost less than 3000 Euro each, can be deployed in any urban environment and their combined experimental information used for the detection of ultra high energy showers heaving an Earth-surface footprint of the order a square kilometer. Moreover, with the deployment of such systems on high schools various educational activities can be created and give students the opportunity to be involved in a modern physics experiment \cite{helycon2}.
\section*{Acknowledgments}
This research was funded by the Hellenic Open University Grant No. $\Phi$K 228: ``Development of technological applications and experimental methods in Particle and Astroparticle Physics''

\end{document}